\documentclass[aps,floatfix,twocolumn,showpacs,preprintnumbers,amsmath,amssymb,a4paper]{revtex4}
\usepackage{latexsym}
\usepackage{amsfonts}
\usepackage{amsmath}
\usepackage{bm}
\usepackage{amssymb}
\usepackage{graphicx}
\usepackage[usenames]{color}

\newcommand{\figref}[1]{\mbox{Fig.~\ref{#1}}}
\newcommand{\tabref}[1]{\mbox{Table~\ref{#1}}}

\newcommand{\eqnref}[1]{\mbox{Eq.~(\ref{#1})}}
\newcommand{\ssman}{{\Sigma}}
\newcommand{\pdfrac}[2]{\frac{\partial #1}{\partial #2}}
\newcommand{\mapF}{{\mathcal F}_{\ssman}}
\newcommand{\tcap}{t_{\mathrm{trap}}}
\newcommand{\tesc}{t_{\mathrm{esc}}}
\def\torus{\mathbb{T}}
\def\reals{{\mathbb R}}
\def\naturals{{\mathbb N}}
\def\integers{{\mathbb Z}}
\def\ocsOa{{\mathcal O}_a}
\def\ocsOc{{\mathcal O}_b}
\def\ocsOf{{\mathcal O}_c}
\def\VMorse{V^{\mathrm{M}}}

\begin{document}

\title{{Bottlenecks to vibrational energy flow in {OCS}: Structures and
    mechanisms}}

\author{R. Pa\v{s}kauskas$^{1}$\footnote{
    Present address: Sincrotrone Trieste,
    AREA Science Park,
    34012 Basovizza Trieste, ITALY}}
\email{rytis@gatech.edu}
\author{C. Chandre$^2$}
\author{T. Uzer$^1$}

\affiliation{$^1$ Center for Nonlinear Sciences, School of Physics,
  Georgia Institute of Technology, Atlanta, GA 30332-0430, U.S.A.\\
  $^2$ Centre de Physique Th\'eorique\footnote{UMR 6207 of the CNRS,
  Aix-Marseille and Sud Toulon-Var Universities. Affiliated with the
  CNRS Research Federation FRUMAM (FR 2291). CEA registered research
  laboratory LRC DSM-06-35.} -- CNRS, Luminy - Case 907, 13288
  Marseille cedex 09, France}

\date{\today}

\begin{abstract}
  Finding the causes for the nonstatistical vibrational energy
  relaxation in the planar carbonyl sulfide (OCS) molecule is a
  longstanding problem in chemical physics: Not only is the relaxation
  incomplete long past the predicted statistical relaxation time, but
  it also consists of a sequence of abrupt transitions between
  long-lived regions of localized energy modes. We report on the phase
  space bottlenecks responsible for this slow and uneven vibrational
  energy flow in this Hamiltonian system with three degrees of
  freedom. They belong to a particular class of two-dimensional
  invariant tori which are organized around elliptic periodic
  orbits. We relate the trapping and transition mechanisms with the
  linear stability of these structures.
\end{abstract}

\pacs{34.30.+h, 34.10.+x, 82.20.Db, 82.20.Nk}

\maketitle

\section{Introduction}

How does vibrational energy travel in molecules? Answering this
question succinctly seems a hopeless task considering the complexity
of interatomic interactions in a molecule. Yet even before scientists
were burdened by this knowledge, the so-called statistical theories
posited the answer: Vibrational energy travels ``very fast'' and
distributes itself statistically among the vibrational modes of a
molecule, assumed to resemble an assembly of coupled oscillators, well
before a reaction takes place. Reaction rate theories based on these
assumptions -- known collectively as statistical or RRKM theories
~\cite{rateproc41,robinson72,forst73,pechukas76} -- remain reliable
working tools of the practicing chemist because they have been
vindicated in an overwhelming number of chemical reactions.

However, numerical studies of Hamiltonian systems have provided solid
evidence~\cite{fpu55,fpu65,ford92,ruffo05,zaslavsky05,carati05} that
the approach to equilibrium usually proceeds more slowly than
predicted by statistical theories~\cite{carter82,uzer91}~--~ and it
is also nonuniform, showing intriguing fits and starts.  In
particular, for Hamiltonian systems with two degrees of freedom, the
familiar picture of chaotic seas, rigid boundaries in terms of noble
tori~\cite{LichtenLieber92}, leaky barriers in terms of cantori~\cite{lobe1,lobe2} has been
well-established in the literature, and these structures are found to
be the source of anomalous transport in such
systems~\cite{zasl02,zasl05}.

Beyond two degrees of freedom, the transport picture in terms of phase space structures is less clear. However, the phase space of higher-dimensional systems shows similar features such as the abundance of periodic orbits, and a mixture of chaotic and regular regions, the latter being characterized (under some hypothesis) by invariant tori of various dimensions. The {KAM} theorem~\cite{LichtenLieber92} states that these structures are in general robust with respect to an increase of the perturbation or equivalently to an increase of energy. Understanding transport properties has to rely on these robust structures which are encountered by any typical trajectory. Roughly speaking, the presence of so many periodic orbits explains why generic trajectories, even when the system is strongly chaotic,
display long intervals of near-regular behavior alternating with fits of chaos--a hallmark of anomalous diffusion. 

The slow approach to equilibrium started to be acknowledged a little
over fifty years ago with the investigation of the dynamics of coupled
oscillators by Fermi, Pasta and Ulam who showed that the relaxation
problem is far more complex than
anticipated~\cite{fpu55,fpu65,ford92,ruffo05,zaslavsky05,carati05}.
In chemical physics, anomalous diffusion was first implicated in the
intramolecular vibrational energy relaxation of the carbonyl sulfide
OCS molecule~\cite{carter82}. The numerical study of a classical
Hamiltonian model of {OCS} shows very slow energy
redistribution among the vibrational modes, even in the fully chaotic
regime~\cite{carter82}, disagreeing strongly with the fast timescales
derived from traditional statistical theory. The understanding of the dynamics was successfully achieved for a collinear model of OCS which has two degrees of freedom~\cite{davis85,davis86,rice87,martens87,skodje88}. However, severe technical difficulties~\cite{gillilan90,ezra,mikitotoda} have prevented such a level of understanding beyond two degrees of freedom, and in particular, for the planar OCS model, in which the molecule is allowed to bend.

In this paper, we analyze the dynamics of a model for the planar OCS
which is a Hamiltonian system with three strongly coupled degrees of
freedom. The aim is to identify the relevant structures in phase space
which are responsible for trappings and escapes, strongly influencing
the transport properties (most prominently, the redistribution of intramolecular energy
among the three modes). For example, rapid diffusion through phase
space takes place through the so-called accelerator
modes~\cite{vered}. In contrast, sticky structures~\cite{conto02} like resonant
islands or tori influence the dynamics by strongly slowing down the
trajectories passing nearby. All these structures are responsible for
anomalous diffusion and fractal kinetics in the system (for recent
surveys, see Refs.~\cite{zasl02,zasl05} and references
therein). Identifying these structures and the mechanisms behind
trapping, escape and roaming is essential for understanding the
transport properties of a given system. Given that there are many such
structures in a realistic system, the only realistic hope for forming
a generally valid picture of transport is to locate invariant
structures which are responsible for the main changes in the transport
properties.

The specific question we address is: What are the structures
in the phase space of OCS acting as dynamical bottlenecks to
the diffusion of chaotic trajectories? What are the structures allowing transitions to other parts of phase space? For three degree of freedom systems, these invariant structures
can be invariant tori with dimensions zero (stagnation points), one (periodic orbits),
two or three~\cite{FA,froe97a,cincotta00}. They can also
include the stable and unstable manifolds of these objects~\cite{wiggins92}. How are invariant structures relevant in the phenomena of capture in chaotic systems? For planar OCS, we find that the bottlenecks and the transition mechanisms from trapped to hyperbolic behavior are provided by a particular class of two-dimensional tori and their unstable manifolds. These results were recently announced in a Letter~\cite{paskauskas08}.   

The paper is organized as follows: In Sec.~\ref{sec:known}, we briefly recall some basics of the Hamiltonian model for the planar rotationless OCS molecule. We also summarize the main results obtained on the dynamics of OCS relevant to the transport properties (both in the planar and collinear cases). In Sec.~\ref{sec:transition}, we illustrate the transitions which occur in the neighborhood of periodic orbits using several representations: Time series,  time-frequency analysis, and Poincar\'e sections. The striking common feature exhibited by many trajectories support the idea of some kind of universal transition mechanism. In Sec.~\ref{sec:phasespace}, after summarizing our methodology, we investigate the neighboring phase space structures which strongly influence the dynamics of these trajectories. 

\section{The OCS model}
\label{sec:known}
\subsection{The Hamiltonian}

\begin{figure}[t]
  \begin{center}
    \includegraphics[width=0.40\textwidth]{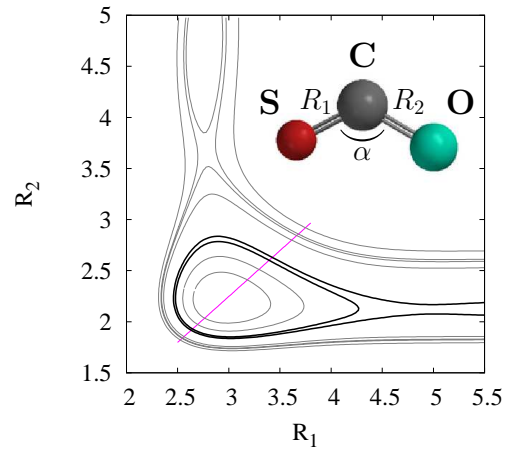}
    \caption{\label{fig:equipotential} 
    Equipotential surfaces of the
      collinear configuration, given by $V(R_1,R_2,\pi)=E$ [see
      Eq.~(\ref{e:U})]. From center outwards, energies are $E=0.03$, $0.06$,
      $0.09$, $0.10$, $1.6$, $1.75$, $1.8$, $2.0$. The energies studied in
      this article are $0.09$ (below dissociation of the weakest bond)
      and $0.10$ (above dissociation), and the corresponding
      equipotential contours are shown in bold.
      }\end{center}
\end{figure}

The dynamics of the planar model of carbonyl sulfide (OCS) can be
described by a Hamiltonian model with three degrees of freedom with
three strongly coupled, non-separable modes: There are two stretching
modes and one bending mode. Each mode is represented by a coordinate
and momentum pair, which we define as: $R_1=d(C,S)$ and $P_1$ for the
CS stretching mode, $R_2=d(C,O)$ and $P_2$ for the CO stretching mode,
and finally, $\alpha=\angle{(OCS)}$ and $P_{\alpha}$ for the bending
mode of the molecule (see~\figref{fig:equipotential}). Hamiltonian model for the rotationless {OCS}
molecule has been provided in~\cite{foord75,carter82}. It has the form
\begin{equation}
  \label{e:ham}
  \begin{split}
    H(R_1,R_2,\alpha,P_1,P_2,P_\alpha)=&\,T(R_1,R_2,\alpha,P_1,P_2,P_\alpha)\\&+V(R_1,R_2,\alpha)\,,
    \end{split}
\end{equation}
where $T$ is the kinetic energy and $V$ is the potential energy. The kinetic
energy is quadratic in the momenta and is provided as
\begin{eqnarray*}
  T&=&\frac{\mu_1}{2}P_1^2+\frac{\mu_2}{2}P_2^2+\mu_3P_1P_2\cos\alpha\\
  && +P_\alpha^2\left(
    \frac{\mu_1}{2R_1^2}+\frac{\mu_2}{2R_2^2}-\frac{\mu_3\cos\alpha }{R_1
      R_2}\right)\\
  &&-\mu_3P_\alpha \sin \alpha \left(\frac{P_1}{R_2}+\frac{P_2}{R_1}\right)\,,
\end{eqnarray*}
where $\mu_i$ are the reduced masses.
Based on available experimental data, the analytic model for the
potential energy surface has been proposed in~\cite{foord75}. In summary, $V$ is given by
\begin{equation}
  \label{e:U}
  V(R_1,R_2,\alpha)=\sum\limits_{i=1}^3 V_i(R_i)+V_I(R_1,R_2,R_3)\,,
\end{equation}
where $V_i(R_i)$ are Morse potentials for each of the three
interatomic distances $R_1$, $R_2$ and $R_3=d(S,O)$, and
\begin{equation}
  V_i(R)=D_i\left(1-\exp{[-\beta_i(R-R_i^0)]}\right)^2\,.
\end{equation}
Here, $R_i^{0}$ are the equilibrium interatomic distances, and $R_3$
is given by $R_3(R_1,R_2,\alpha)=(R_1^2+R_2^2-2R_1R_2\cos{\alpha})^{1/2}$. 
At equilibrium, the molecule is collinear, therefore
$R_3^{0}=R_1^{0}+R_2^{0}$.
Also, the interaction potential $V_I$ assumes the Sorbie-Murrell form:
$$V_I=A\,P(R_1,R_2,R_3)\prod_{i=1}^3\left(1-\tanh\gamma_i[ R_i-R_i^0]\right)\,,$$
where $P(R_1,R_2,R_3)$ is a quartic polynomial in each of its variables:
\begin{equation*}
  \begin{split}
  P(R_1,R_2,R_3)=1&+c_{i}^{(1)}R_i+c_{ij}^{(2)}R_iR_j\\
   &+c_{ijk}^{(3)}R_iR_jR_k+c_{ijkl}^{(4)}R_iR_jR_kR_l\,.
  \end{split}
\end{equation*}
All the coefficients ($\mu_i$, $D_i$, $\beta_i$, $R_i^{0}$,
$\gamma_i$, $A$, $c_{i}^{(1)}$, $c_{ij}^{(2)}$, $c_{ijk}^{(3)}$,
$c_{ijkl}^{(4)}$) are provided in Ref.~\cite{carter82}.  We display
the equipotential surfaces of $V(R_1,R_2,R_1+R_2)$ of the collinear
configuration in Fig.~\ref{fig:equipotential}.  The equations of
motion can be derived from Hamiltonian~(\ref{e:ham}) using the canonical
Hamilton's equations.

\subsection{Summary of prior results on the OCS dynamics}

The classical models of both the collinear and the planar
(rotationless) carbonyl sulfide OCS molecule have been studied in
detail in
Refs.~\cite{carter82,davis84,dawa84,davis85,ezra2,wiggins92,martens87}.

The dynamics in the collinear configuration of {OCS} was first studied
by Carter and Brumer~\cite{carter82}. They characterized the motion of
this system at a number of energies, extending up to
$20,000\,\mathrm{cm}^{-1}$ (which amounts to $E=0.09$ a.u.)  A
relaxation time, as defined in
Refs.~\cite{sinai73,farantos81,hamilton83}, was estimated at $0.17$
pico-seconds. However, after integrating trajectories for $2.4$
picoseconds, no relaxation to statistical equilibrium was 
observed. When this contradiction was investigated by integrating the
equations for much longer times (up to $45$ picoseconds), two
distinct timescales for relaxation were found, the longer of which
characterized energy redistribution that was incomplete even after
$45$ picoseconds~\cite{dawa84}. Even on the picosecond time scale,
sudden transitions between relatively long-lived regions of localized
mode energies were observed.  Since this collinear model has two
degrees of freedom, Davis and Wagner~\cite{dawa84} used Poincar\'{e}
surfaces of section as a visualizing tool for phase space
structures. These revealed that even at high energy ($E=0.09$), the
system has a ``divided phase space'', with coexisting regular and
chaotic regions. They observed that trajectories can be trapped in
restricted regions of phase space for many vibrational periods, after
which they would suddenly move to other regions of phase space to
repeat the pattern.

Progress came with the recognition that the then-recent lobe
dynamics~\cite{lobe1,lobe2} could help to explain non-statistical
relaxation in two degree of freedom systems~\cite{davis85}. When the
strength of the perturbation (or equivalently, the total energy) is
increased, the two-dimensional invariant tori of a Hamiltonian system
with two degrees of freedom develop sets of ``holes'' with the
systematics of Cantor sets. These holes, dubbed
``cantori''~\cite{lobe1}, form leaky barriers which can act as
bottlenecks to phase space transport. These bottlenecks are associated
with broken tori with irrational frequency ratios, where those with
``noble'' number ratios being generically the very last to be
destroyed by an increasing perturbation (the supporting argument being
that these numbers are the most poorly approximated by
rationals~\cite{hardy79}).  For OCS, their existence has been
confirmed in Ref.~\cite{davis85} in a region between two resonances
$\omega_{\mathrm{CO}}/\omega_{\mathrm{CS}}=3/1$ and
$\omega_{\mathrm{CO}}/\omega_{\mathrm{CS}}=5/2$.  The noblest
irrational number between the rationals $5/2$ and $3/1$ is $2+\gamma$,
where $\gamma=(\sqrt{5}-1)/2$ is the golden
mean~\cite{lobe1,hardy79} and can be expressed as a continued fraction
of an infinite sequence of ones, also written as $[1,1,1,1,\ldots]$.
These results obtained from classical mechanical were confirmed using
quantal wave packet calculation~\cite{gibson86}. However, these
successful results could not be extended to the planar {OCS} due to
severe technical and computational
difficulties~\cite{gillilan90,ezra}. Yet there were indications that
this problem of intramolecular energy flow in higher dimensions is
also related to the resonant and non-resonant
structures~\cite{martens87,ezra2}. In particular, the relevance of
Arnold's web in the diffusion of trajectories was highlighted. Among
their conclusions are that transport is most rapid along low order
resonance zones; transport is slow (diffusive) along high order
resonances; it was conjectured that pairwise noble frequency ratios
play a role of inhibiting transport along resonance lines.

\section{Trappings and transitions in the planar OCS: bottlenecks and transition mechanisms}

\subsection{Observations}
\label{sec:transition}
The complexity of transport processes in the collinear {OCS} model,
revealed in the early investigations, suggests that a look into phase
space structures such as periodic orbits or invariant tori is needed
for a better understanding of these processes. Even if the measure of
such invariant structures embedded into a chaotic sea is typically
zero, the ``neighborhoods'' of influence around them can have
relatively large measure and their finite-time properties, as
characterized by Lyapunov exponents, provide a quantitative picture of
transport. The rationale goes as follows~: An ensemble of trajectories,
described by a density function, which is centered in a finite volume
around a periodic orbit, will evolve in finite time following this
periodic orbit, and spreading predominantly in the direction of
unstable manifolds, exponentially in time with a rate equal to the local Lyapunov exponent.  An orbit in the ``neighborhood'' of a periodic
orbit, temporarily assumes or ``shadows'' the properties of this
periodic orbit as a general consequence of dynamical
continuity~\cite{lochak}. This temporary influence of
periodic orbits can also be viewed as instantaneous time-periodic
forcing, exerted by a periodic orbit. It is expected that, in general,
typical trajectories are trapped for longer times in the neighborhoods
of linearly stable orbits. In what follows, we draw a dynamical
picture of transport in OCS based on the determination of invariant
structures in phase space and their linear stability properties.

\subsubsection{Density of periodic orbits}
\begin{figure}
  \begin{center}
    \includegraphics[width=8.5cm]{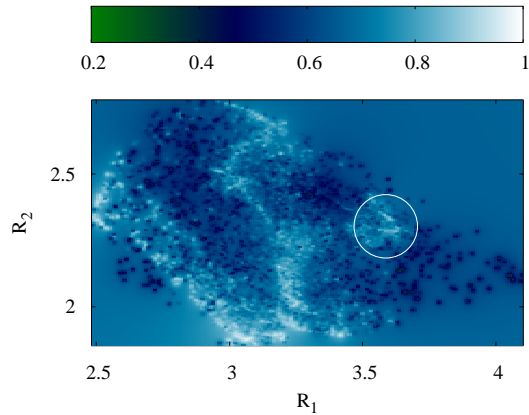}
    \caption{ 
      \label{fig:density_po}
      Averaged density of periodic orbit points on the Poincar\'e surface of
      section, projected onto the $(R_1,R_2)$-plane, weighted by the
      ``local escape rate'' $\gamma^{+}_p$, the sum of
      positive Lyapunov exponents, $\lambda^{(p)}_i>0$ or, in terms of
      Lyapunov multipliers for the periodic orbit $p$, given by
      $\gamma^{+}_p=\prod_{i: \lvert\Lambda^{(p)}_i\rvert \geq 1}
      \lvert \Lambda^{(p)}_i\rvert^{-1/T_p}$ where $\Lambda_i^{(p)}$
      is an eigenvalue of $D\mapF$ evaluated at the periodic points.
      Periodic orbits with the following number of returns to the
      Poincar\'e sections are determined: $1(4)$, $2(9)$, $3(10)$,
      $5(24)$, $7(26)$, $8(101)$, $11(40)$, $13(33)$, $17(21)$,
      $19(43)$, $23(41)$, $29(34)$, $31(28)$, $37(43)$ where the
      number of orbits is shown in parentheses. Energy is set at
      $E=0.09$. Lighter areas are dominated by more regular orbits,
      darker by unstable orbits. The circle indicates the region
      located near $\ocsOa$ where $(R_1,R_2)\approx (3.6,2.3)$ where
      trapping and roaming is analyzed in Fig.~\ref{fig:tfa1}.}
  \end{center}
\end{figure}

\begin{figure}[t]
  \includegraphics[width=8.5cm]{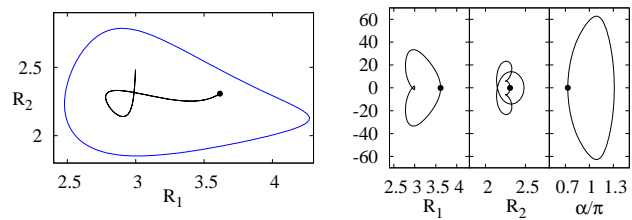}
  \caption{\label{fig:ocsOa}
    The periodic orbit $\ocsOa$ at $E=0.09$: projections in the
    $(R_1,R_2)$-plane (left panel) and in the $(R_1,P_1)$, $(R_2,P_2)$
    and $(\alpha,P_\alpha)$ planes (right panels). The blue curve in
    the $(R_1,R_2)$ projection is the boundary of the energetically
    accessible region. Dots indicate the location of the
    intersection with the Poincar\'e section $\ssman$. The
    periodic orbit $\ocsOa$ is of elliptic-elliptic linear stability
    type (see Tab.~\ref{tab:crd1} for details).}
\end{figure}

A generic feature of Hamiltonian dynamics is the abundance of periodic
orbits in phase space. Figure~\ref{fig:density_po} represents the
averaged density in the configuration space $(R_1,R_2)$ of periodic
orbit points on the Poincar\'e section $\ssman$ (defined in
Sec.~\ref{sec:poinc}) for planar OCS. A closer inspection of this figure
shows that the most prominent regions of stability surround short
periodic orbits with elliptic linear stability.
A typical trajectory passes through this maze of periodic orbits,
being trapped for some time according to local stability
properties. The aim of this manuscript is to understand how a typical
trajectory can be trapped and released locally around a given periodic
orbit. In what follows we analyze the transport properties in the
neighborhood of an elliptic periodic orbit, like for instance $\ocsOa$, as shown by a circle
in~\figref{fig:ocsOa}.

\begin{figure}
  \includegraphics[width=8.5cm]{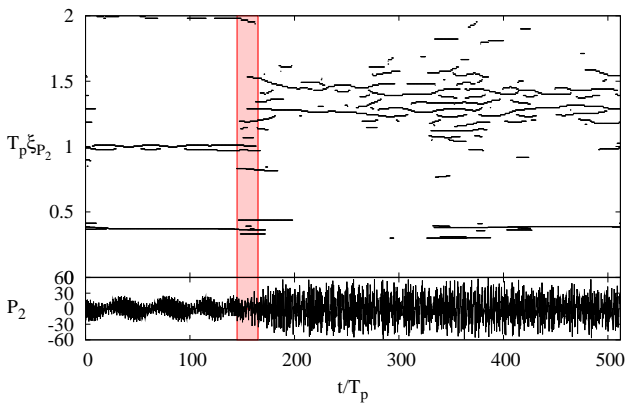}
  \caption{\label{fig:tfa1} Lower panel: Time series $P_2(t)$, of a
    trajectory with initial coordinate $\sigma_1(\ocsOa)$ (see
    Tab.~\ref{tab:crd1}) near the periodic orbit $\ocsOa$. The energy is $E=0.09$.  Time is scaled to
    $T_p^{(\ocsOa)}$, the period of $\ocsOa$. The
    integration time is $T_{\text{max}}=512T_p^{(\ocsOa)}\approx
    34\,\text{ps}$.  Upper panel: Ridges of the time-frequency decomposition of $P_2(t)$. The
    frequencies of $P_2(t)$ are denoted $\xi_{P_2}$, and are
    represented in units of $(T_p^{(\ocsOa)})^{-1}$. The shaded band
    locates the transition region.}
\end{figure}

\begin{figure}[t]
\includegraphics[width=8.5cm]{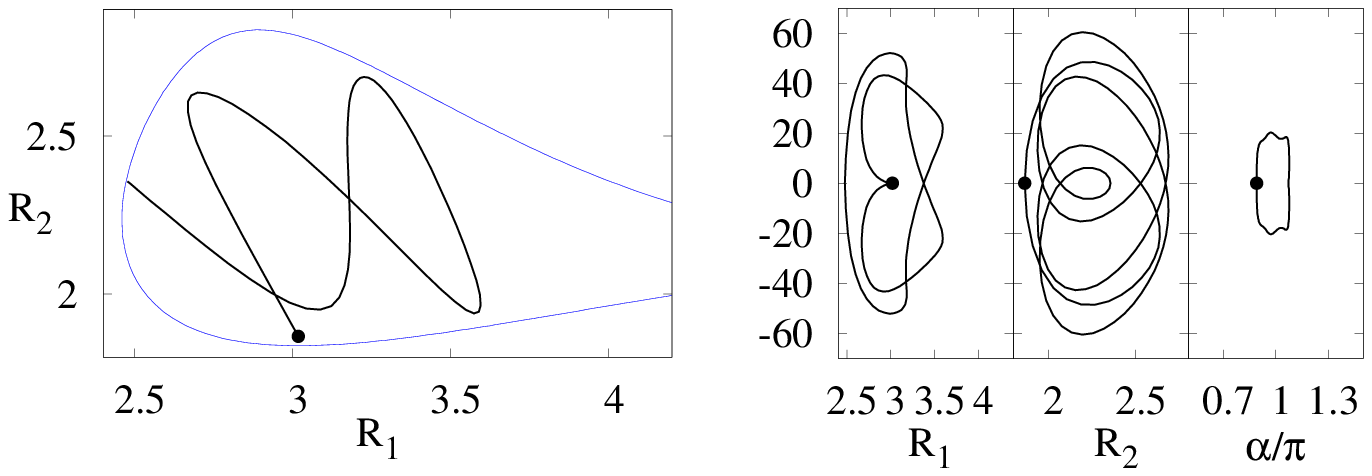}
  \includegraphics[width=8.5cm]{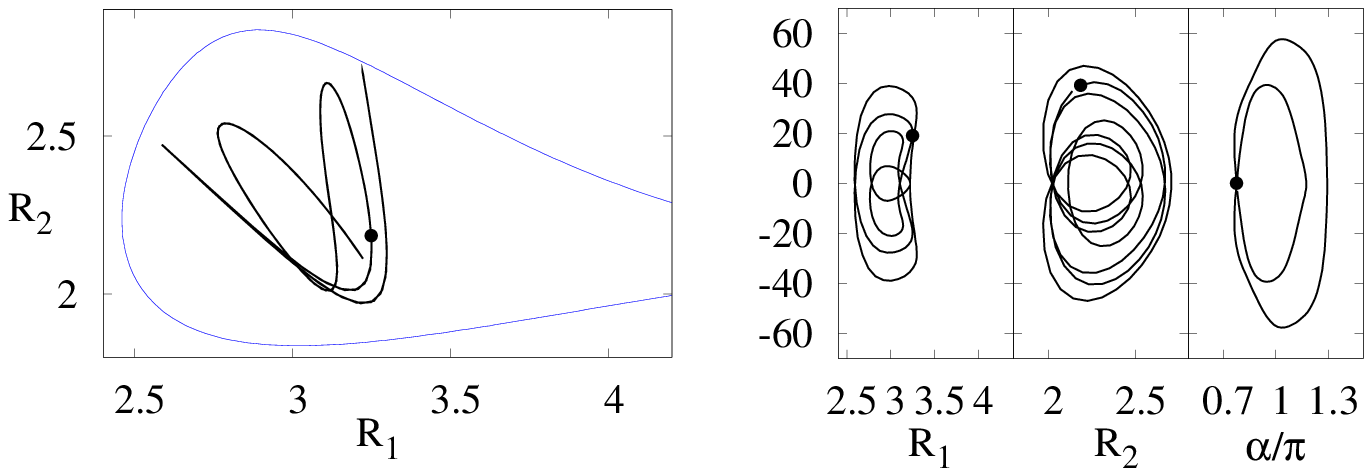}
  \caption{\label{fig:ocs_OcOf} Two stable periodic orbits, $\ocsOc$
    (upper panel) and $\ocsOf$ (lower panel) for $E=0.1$: projections
    in the $(R_1,R_2)$-plane (left panel) and in the $(R_1,P_1)$,
    $(R_2,P_2)$ and $(\alpha,P_\alpha)$ planes (right panels). These
    orbits are relevant in the trapping of trajectories with initial
    conditions $\sigma_2(\ocsOc)$ (see Tab.~\ref{tab:crd1}). The zero
    velocity curve (boundary in ($R_1$, $R_2$) coordinates) is shown
    in blue. Dots indicate the location of the
    intersection with the Poincar\'e section $\ssman$.}
\end{figure}

\begin{table}[ht]
  \begin{tabular}{|c|r@{.}l|r@{.}l|r@{.}l|}
    \hline
$\ocsOa$ &3&6151934418418414&     0&0000000000000\\
$0.09$   &2&3075960024093884&     0&0000000000000\\
         &2&2933715073362912&     0&0000000000000\\
         \hline
$T_p^{(\ocsOa)}$      &\multicolumn{4}{l|}{2622.68398495968 {\rm a.u.}\, (0.06344 {\rm ps})}  \\
         \hline
$\omega^{(\ocsOa)}/\pi$\footnote{Note that in Ref.~\cite{paskauskas08} the second return map was considered so the stability indices of $\ocsOa$ are half of the ones here.}&  0&4900126616&   0&7409374404 \\
\hline
\hline
$\ocsOc$       &3&01777791580821&    0&0000000000000\\
 0.10  &1&86626520637548&       0&0000000000000\\
       &2&79762936317876&       0&0000000000000\\
       \hline
$T_p^{(\ocsOc)}$    &\multicolumn{4}{l|}{3662.61014507030904 {\rm a.u.} \, (0.08859 {\rm ps})}\\
       \hline
$\omega^{(\ocsOc)}/\pi$&  0&4560162021&   0&3224799075 \\
       \hline
       \hline
$\ocsOf$       &3&24837693124009&  19&09854653413159\\
 0.10  &    2&18475382468168&  39&13951947510928\\    
       &    2&43942803794946&   0&0\\
       \hline
$T_p^{(\ocsOf)}$    &\multicolumn{4}{l|}{5119.91498417653838 {\rm a.u.} \, (0.12385 {\rm ps})}\\
\hline
$\omega^{(\ocsOf)}/\pi$& 0&7083411883&   0&1546899545\\
       \hline
\hline
$\sigma_1(\ocsOa)$      &3&63724286026980&    -0&00109450235083\\
$0.09$   &2&25801058566880&     0&29106367555962\\
         &2&29583595102985&     0&00000000000000\\ 
       \hline
       \hline
$\sigma_2(\ocsOc)$    &3&05250153680800&     0&01979520051700\\
$0.10$   &1&84284317381100&    -0&10285514011200\\
         &2&98055080176000&     0&00000000000000\\
       \hline
  \end{tabular}
  \caption{\label{tab:crd1}
    Initial conditions of the trajectories considered in the manuscript: The three periodic orbits $\ocsOa$, $\ocsOc$ and $\ocsOf$, and the two trajectories $\sigma_1(\ocsOa)$ and $\sigma_2(\ocsOc)$, one close to $\ocsOa$ and the second one to $\ocsOc$. 
    First column: label of the initial conditions, value of energy $E$, and in case of periodic orbit, period $T_p$ and rotation numbers $\omega$. Second column: $R_1$, $R_2$ and $\alpha$. Third column: $P_1$, $P_2$, and $P_\alpha$. 
  }
\end{table}

\subsubsection{Time-frequency analysis and stroboscopic mapping}\label{s:tfa}
\begin{figure}
  \includegraphics[width=8.5cm]{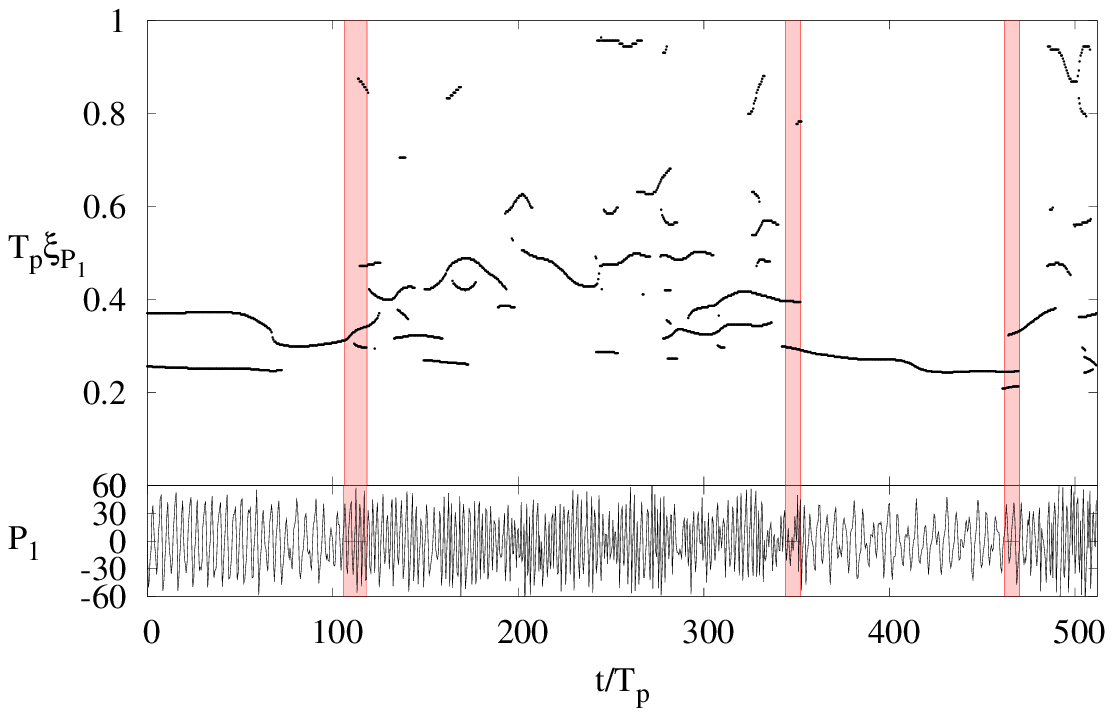}
  \caption{\label{fig:tfa3} Lower panel: Time series $P_1(t)$ of a
    trajectory initially close to $\ocsOc$ with initial coordinates
    $\sigma_2(\ocsOc)$ (see Tab.~\ref{tab:crd1}). The energy is $E=0.1$. The
    integration time is approximately $512T_p^{(\ocsOc)}$.  Upper
    panel: Ridges of the time-frequency decomposition of $P_1(t)$. We
    notice that they are three transition regions (located by shaded
    bands) in this case.}
\end{figure}

To examine the temporal features of trajectories we use time-frequency
analysis~\cite{chandre03}. In what follows, ${\bf x}$ designates a
point in phase space, i.e.\ ${\bf
x}=(R_1,R_2,\alpha,P_1,P_2,P_\alpha)$. A finite segment of a
trajectory can be represented by a sequence of phase space
coordinates, $\{\mathbf{x}_n\}_{n=1,\ldots,N}$,
$\mathbf{x}_n=\mathbf{x}(t_n)$, visited by a trajectory at times
$t_n$. For a stroboscopic map, we take snapshots with a fixed time
increment, $t_{n+1}=t_n+\Delta$. It is natural to scale the time
increment $\Delta$ by the period $T_p$ of the organizing periodic
orbit. We select $\Delta=T_p/4$. The time series of selected orbits
are displayed in the bottom panels of Figs.~\ref{fig:tfa1}
and~\ref{fig:tfa3}.

We study the instantaneous frequencies using wavelet decomposition. As
described in Refs.~\cite{chandre03,tf96,carm98}, the time-frequency
analysis is based on a continuous wavelet transform of an observable
$f(t)$

\begin{equation}
  \label{e:cwt} W
  f(t,s)=\frac{1}{\sqrt{s}}\int_{-\infty}^{+\infty} f(\tau) \psi^*
  \left( \frac{\tau-t}{s}\right) d\tau\,.
\end{equation}
We choose the mother wavelet $\psi$, in the Morlet-Grossman form:
$\psi(t)= e^{\iota \eta t}e^{-t^2/2\sigma^2}/(\sigma^2 \pi)^{1/4}$,
with adjustable parameters $\eta$ and $\sigma$. The time-frequency
representation is obtained via a relation between the scale $s$ and
the frequency $\xi=\eta/s$. We consider the normalized scalogram
$$ P_W f(t,\xi=\eta/s)=\vert W f(t,s)\vert^2/s\,,$$ which can be
interpreted as the energy density in the time-frequency plane.  The
ridges of $P_W$ can be interpreted as instantaneous frequencies, or
more rigorously, the set of frequencies for a given time interval.  In
this section, two typical trajectories (whose initial conditions are
specified in Tab.~\ref{tab:crd1}) which are initially close to
elementary organizing periodic orbits, are represented in
Figs.~\ref{fig:tfa1} and \ref{fig:tfa3} where the signal $f(t)$ is
chosen to be $P_1(t)$ or $P_2(t)$. It should be noticed that other
choices of observables $f(t)$ lead to the same qualitative features as
the ones presented here, and in addition, these features are common to
a wide set of other trajectories in the same neighborhood.

Time-frequency analysis shows that each of these trajectories displays
qualitatively distinct regions~: some with approximately constant ridges in time, and
others with multiple and short ridges. These two regimes are clearly
marked with transition intervals (highlighted by shaded bands in the
figures). In Fig.~\ref{fig:tfa1} we observe a clear, sharp transition
stage between trapped behavior (around $\ocsOa$) and roaming behavior
throughout a large portion of phase space. After some time spent
around the periodic orbit, the trajectory seems to find an exit
channel through a bottleneck. Generically, any trajectory experiences
multiple events of capture and escape (like the one in
Fig.~\ref{fig:tfa3}).  We have found that escape to the chaotic region
proceeds in two stages, characterized by two different rates of
escape. The transition interval is characterized by $\tcap$ and
$\tesc$.  The first ({\it ``slow''}) stage, $0\leq t\leq\tcap$, and
the second ({\it ``fast''}) stage $\tcap\leq t\leq \tesc$. The precise
transition points located at $\tcap$ and $\tesc$ may vary in different
situations, but typically $\tesc-\tcap \ll \tcap$.  We notice that
these trapping and transition stages although visible, were not as
clearly apparent on the time series as on the time-frequency plots.

In order to identify the phase space regions visited during the
trapping and escape stages, we complement the time-frequency analysis
by projections of segments of the time series in a two-dimensional
plane given by two coordinates, e.g., $(R_1,R_2)$, $(R_1,P_1)$ or
$(R_2,P_2)$. In Fig.~\ref{fig:str1}, two segments of the trajectories
of Fig.~\ref{fig:tfa1} are represented (left and middle panels), one
corresponding to the trapped stage (to the left of the shaded band),
and the other one to the trajectory after the escape process (to the
right of the shaded band). It is shown that the trajectory is trapped
into a small L-shaped region around the periodic orbit $\ocsOa$, and
that after the transition point, the trajectory has access to a larger
part of phase space with an apparent size of the order of the entire
accessible region. The same observation follows for the trajectory of
Fig.~\ref{fig:tfa3}, the stroboscopic plot of which appears in
Fig.~\ref{fig:str3}. By drawing tubes around it, we notice that the
trajectory in both trapped segments sticks to particular regions
around different periodic orbits.

\subsubsection{Poincar\'{e} sections}\label{sec:poinc}
We use Poincar\'e sections as another way to visualize
multidimensional trajectories. Given that this Hamiltonian system has
three degrees of freedom, the Poincar\'e section is four
dimensional. We show below how two-dimensional projections of these
sections can be used to gain insight into the dynamics (although this
information is displayed less clearly for this system than for a
system with two degrees of freedom). Given some scalar function $U({\bf x})$ of the
phase space variables, we define this section $\ssman$ to
be the set of points $\mathbf{x}$ of a trajectory such that
\[
U(\mathbf{x})=0\,,
\]
with $\dot{\mathbf x}\cdot\partial U/\partial {\bf x} >0$.  From two consecutive points
${\bf x}_n={\bf x}(t_n)$ and ${\bf x}_{n+1}={\bf x}(t_n+\Delta({\bf
x}_n,t_n))$ on the Poincar\'e section, we define a \emph{Poincar\'{e}
map} $\mapF$,
\[ 
\mapF({\bf x}_n) = {\bf x}_{n+1}.
\]
In what follows, we have used the surface $\ssman$ defined by
\begin{equation}\label{e:section}
  U(\mathbf{x}) = P_{\alpha}\,.
\end{equation}
The argument for choosing this surface goes as follows: We study an
energy range where the time series of the bending mode
$(\alpha,P_{\alpha})$ oscillate around an instantaneous mean value
$\langle\alpha\rangle$. Between each oscillation there is a turning
point where momentum $P_{\alpha}$ vanishes.  The only case which is
not captured by the Poincar\'e section is when the bending mode is
``frozen'' to $\alpha=\pi$ and $P_{\alpha}\equiv 0$, which corresponds
to the collinear {OCS}.

We choose a four-dimensional parametrization of the surface of section
$\ssman$ which consists of $R_1$, $P_1$, $R_2$ and $P_2$. Setting
$P_{\alpha}=0$ in Eq.~(\ref{e:HE}), the equation
\begin{equation}
  H(R_1,R_2,\alpha,P_1,P_2,0)=E\,,
\label{e:HE}
\end{equation}
with a constraint $\dot{P_\alpha}>0$ is to be solved for
$\alpha(R_1,P_1,R_2,P_2;E)$ numerically.  There are two merits in
using Poincar\'e sections: First, representing projections as a set of
planar plots of canonically conjugated variables helps in perceiving
the symplectic symmetry of structures.  Second, the section manifold
is one dimension smaller than the energy manifold, and so are the maps
of all the invariant structures. For instance, periodic orbits
correspond to a finite set of points $\{\mathbf{x}_n\}_{n=1,\ldots,N}$
on $\ssman$, and the dynamics visits these points in a cyclic manner,
i.e.\ $\mapF(\mathbf{x}_n) = \mathbf{x}_{n+1}$ for $n=1,\ldots,N-1$,
with $\mapF(\mathbf{x}_N)=\mathbf{x}_1$. Similarly two-dimensional
tori correspond to closed curves on $\ssman$.

\begin{figure*}
\includegraphics*{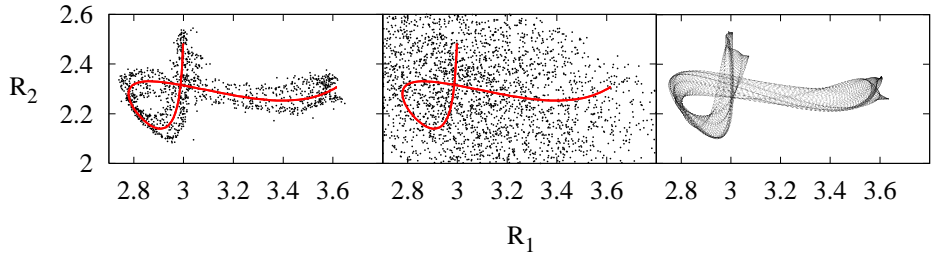}
\caption {Left and middle panels: Stroboscopic plots of segments of
the trajectory of Fig.~\ref{fig:tfa1} (for $E=0.09$): Before the
transition stage (left panel), and after (middle panel). Note how
during the trapping (left panel) the chaotic orbit remains around the
L-shaped stable periodic orbit $\ocsOa$ (represented in red) before
escaping into the chaotic zone (middle panel). The trapping stage is
inside the invariant structure (a two-dimensional torus) shown on the
right panel.}
\label{fig:str1}
\end{figure*}   

\begin{figure*}
  \includegraphics*{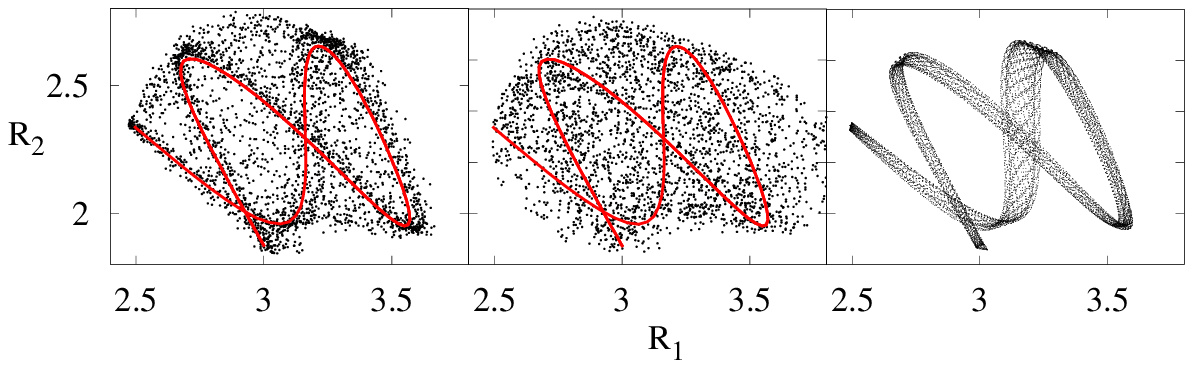}
  \includegraphics*{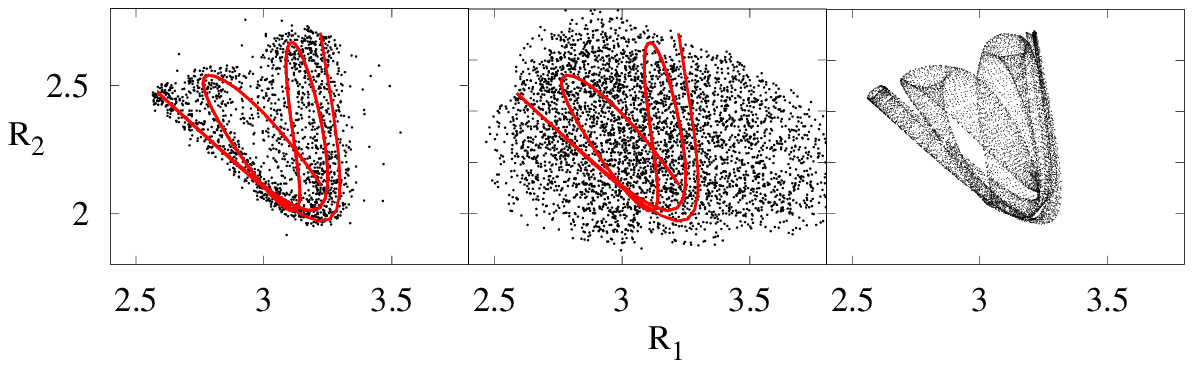}
  \caption{\label{fig:str3} Left and middle panels: Stroboscopic plots of segments of
    the trajectory of Fig.~\ref{fig:tfa3} (for $E=0.1$): trapped stage for
    $t\in[0,110]$ (upper left), chaotic region for $t\in [110,350]$
    (upper middle), trapped stage for $t\in [350,460]$ (lower left),
    and chaotic region for $t\geq 460$ (lower middle). The trapping
    stages (around periodic orbits represented in red) are occurring
    inside the invariant structures (two-dimensional tori) shown on the
    right panels.}\label{fig:str2}
\end{figure*}

\begin{figure*}
  \begin{center}
    \includegraphics[width=6.0cm]{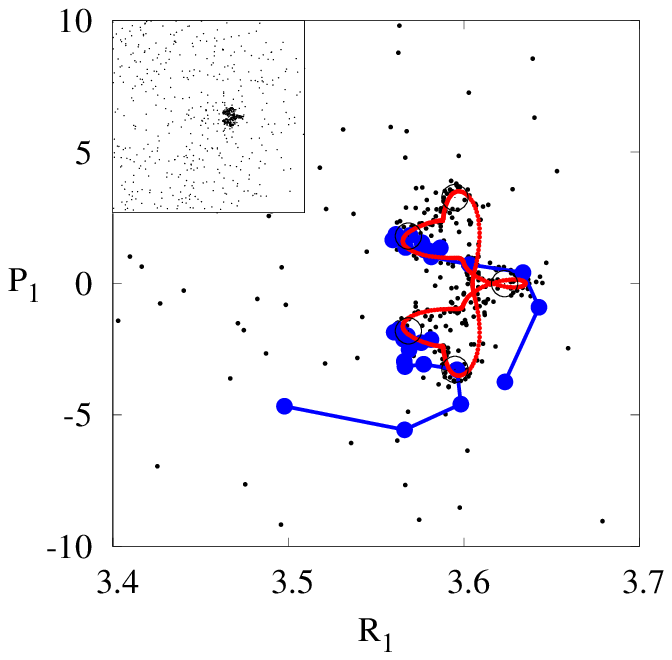}
    \includegraphics[width=6.0cm]{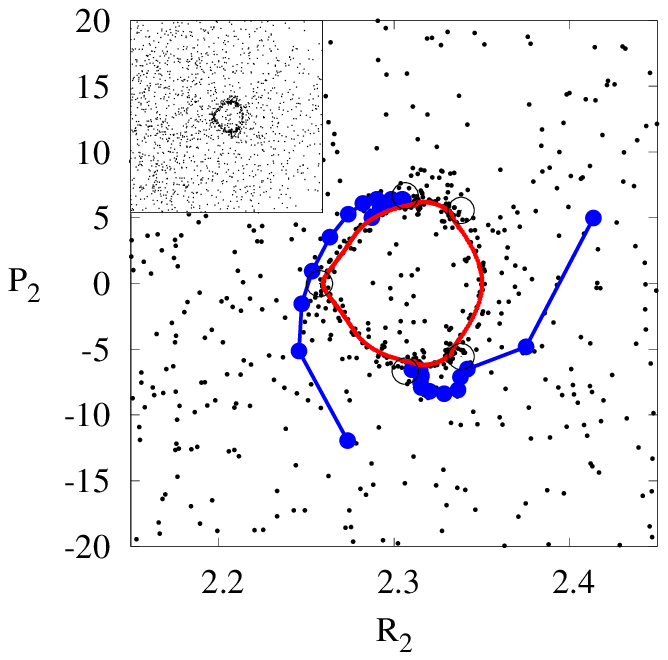}
    \caption{Poincar\'e sections of the trajectory analyzed in
      Fig.~\ref{fig:tfa1} (for $E=0.09$) on the $(R_{1}, P_{1})$-plane (left panel) and on
      the $(R_2,P_2)$-plane (right panel). It is apparent (in the insets)
      that before escaping to the external region, the trajectory is stuck
      in the neighborhood of a well-localized structure. The five points (in
      bold) are the intersections with $\ssman$ of a partially hyperbolic
      resonant periodic orbit which is responsible for the escape to the
      chaotic region through its unstable manifold. Two ``tentacles''
      starting at the upper and lower parts of the structure are marked with
      broken lines connecting crosses to clarify what happens during the
      escape stage (shaded band in Fig.~\ref{fig:tfa1}). The resonant
      2:5 periodic orbit has elliptic-hyperbolic stability with
      $\lambda=0.113231998$ per return (or $\lambda=0.566159992$ for
      the entire orbit) and a rotation number
      $\omega/\pi=0.35684077865194591$ per entire orbit.}
    \label{fig:ps1}\end{center}
\end{figure*}

\begin{figure*}
  \begin{center}
    \includegraphics[width=6.0cm]{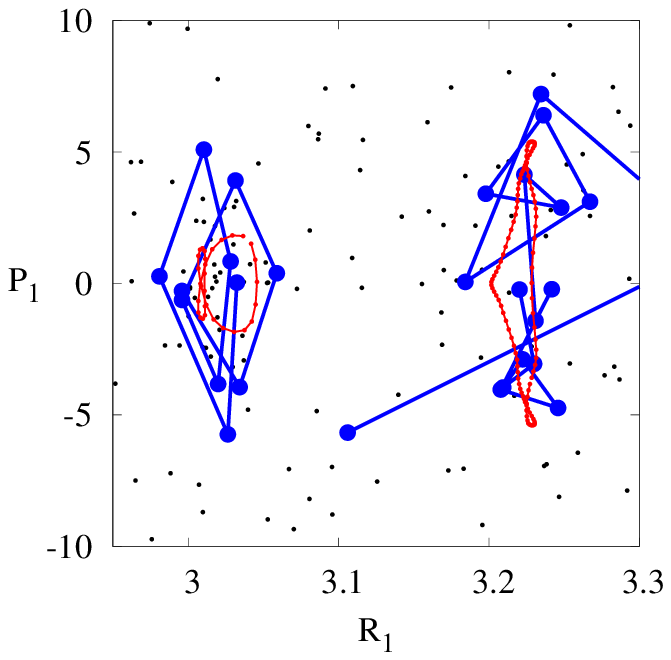}
    \includegraphics[width=6.0cm]{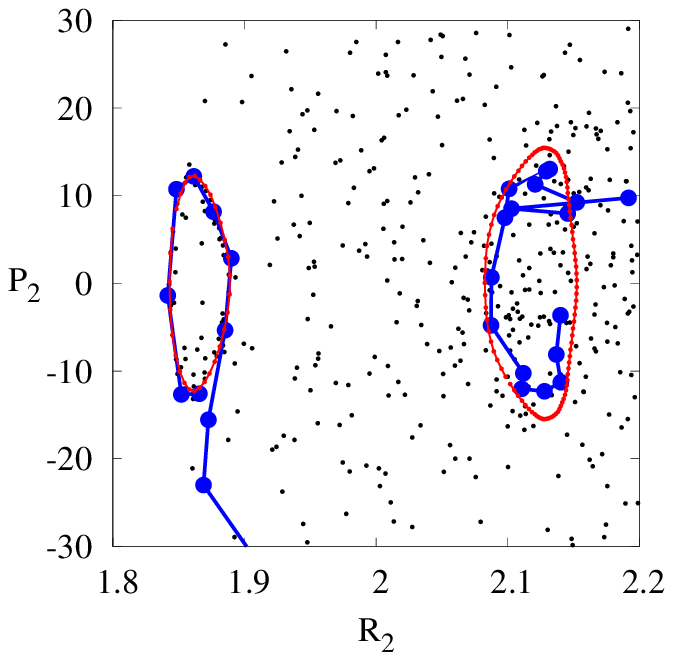}
    \caption{\label{fig:ps3}This figure illustrates the multiple
      capturing which is found generically for a randomly selected
      trajectory, as the one of Fig.~\ref{fig:tfa3} (for $E=0.1$). The quasi-regular
      intervals are color coded: between iterations 1--47 (red) and
      370--430 (blue). Note that the second interval draws two curves,
      because it takes two returns to the surface of section to draw this
      curve (i.e.\ it is an apparently connected curve for $\mapF^2$).
      The two closed loops are the bottleneck torus of the bottom-right panel of
      Fig.~\ref{fig:str2}.}
  \end{center}
\end{figure*}

In Figs.~\ref{fig:ps1} and \ref{fig:ps3}, Poincar\'e sections,
projected on the planes $(R_1,P_1)$ and $(R_2,P_2)$, are drawn for the
two trajectories considered in Figs.~\ref{fig:tfa1} and
\ref{fig:tfa3}. These Poincar\'e sections clearly show distinct
one-dimensional curves (clearly visible in the insets of
Figs.~\ref{fig:ps1} and \ref{fig:ps3}) in the transition stages
(shaded bands on Figs.~\ref{fig:tfa1} and \ref{fig:tfa3}). The tubes
which we identify as two-dimensional invariant tori in phase
space~\cite{paskauskas08}, represented on Figs.~\ref{fig:str1} and
\ref{fig:str3}, correspond to these one-dimensional curves (or more
generally a set of one-dimensional curves) on the Poincar\'e
sections. 

In the trapping stages (around specific periodic orbits) like the
ones in Figs.~\ref{fig:str1} and \ref{fig:str3} (left panels), the
rotation numbers are obtained from the frequency map
analysis~\cite{laskar93} on the surface of section
$\ssman$. Dimensionless ratios of frequencies arise naturally in the
Poincar\'{e} map $\mapF$, and ratios of frequencies are called
\emph{rotation numbers}. The trapping stage can be characterized
by a single rotation number (and its harmonics), implying that a
two-dimensional torus is the relevant invariant structure in the
trapping process. In the following, we determine such structures and
highlight the family of two-dimensional tori which are relevant for
the transport picture in this system.

\subsection{Lower dimensional invariant tori}
\label{sec:phasespace}

\subsubsection{A summary of the methodology}
\label{sec:method}
It is well established that invariant structures in phase space play
an important role in the transport properties associated with
Hamiltonian systems with two degrees of
freedom~\cite{LichtenLieber92}. In particular, the role of periodic
orbits has been singled out in many experiments~\cite{gutbook}. Even
if some aspects of this dynamical picture can be extended to systems
with a larger number of degrees of freedom, it remains to address the
role of invariant structures which are not present in systems with one
and two degrees of freedom, but are specific to three and more degrees
of freedom. In three degree of freedom systems this new type of
invariant structures takes the form of two-dimensional invariant
tori. Observations described in~Sec.~\ref{sec:transition} indicate
that such tori close to elliptic periodic orbits play an important
role. To have a qualitative description of dynamics near a periodic
orbit, we consider a fixed point ${\bf x}_0$ on the surface of section
$\ssman$, i.e.\ $\mapF({\bf x}_0)={\bf x}_0$, corresponding to a point of a
periodic orbit.  Near ${\bf x}_0$, the Poincar\'{e} map
$\mapF$ can be expanded into a linear part and a remainder:
\begin{equation}\label{e:lin}
  \mapF ({\bf x}) = \mapF ({\bf x}_0) + D\mapF({\bf x}_0) ({\bf
  x}-{\bf x}_0) + \mathcal{R}({\bf x}-{\bf x}_0)\,,
\end{equation}
where $D\mapF({\bf x}_0)$ is the matrix of first order derivatives of
the Poincar\'{e} map, constrained to the surface of section and
evaluated at ${\bf x}_0$. All higher order terms in ${\bf x-x}_0$ are
collected in $\mathcal{R}({\bf x}-{\bf x}_0)$.
Finite-time dynamics near the fixed point ${\bf x}_0$ are determined
by the properties of the matrix $D\mapF({\bf x}_0)$. Assuming that
linearized approximation is effective, and discarding the remainder
term from further discussions (the fully nonlinear problem with large
$\mathcal{R}$ is solved using the methodology outlined in
Appendix~\ref{sec:appB}), we consider a closed curve ${\bm \gamma}(s)$
on the Poincar\'e section $\ssman$ defined on a torus $s\in \torus^1$,
and consider the dynamics of ${\bf x}(s) = {\bf x}_0 + \epsilon {\bm
\gamma} (s)$ given by
\begin{equation}\label{e:linnf}
  \mapF({\bf x}(s)) = {\bf x}_0 + \epsilon D\mapF({\bf x}_0){\bm \gamma} (s).
\end{equation}
If $\mapF$ has at least one pair of eigenvalues in the form
$\Lambda=\exp{(\pm\iota\omega)}$, it is possible to find a
${\bm\gamma}(s)$ such that
$D\mapF{\bm \gamma}(s)={\bm\gamma}(s+\omega_{\epsilon})$
and
$\lvert\omega-\omega_{\epsilon}\rvert=o(\epsilon)$.
Therefore the equation 
\begin{equation}
  \label{e:tori-eqn}
  \mapF( {\bf x}(s)) = {\bf x}(s + \omega),
\end{equation}
has a family of solutions, parametrized by the rotation number
$\omega$.  Equation~(\ref{e:tori-eqn}) defines a torus as a loop on
the surface of section $\ssman$ with rotation number $\omega$.  Even
if $D\mapF({\bf x}_0$) has two pairs of eigenvalues of the form
$\exp{(\pm\iota\omega_i)}$ such an invariant loop close to ${\bf x}_0$
can be found. More details on the determination of two-dimensional
invariant tori are given in Appendix~\ref{sec:appB}.

\subsubsection{Invariant tori and their bifurcations: Bottlenecks}
\begin{figure}[t]
  \includegraphics[width=8.5cm]{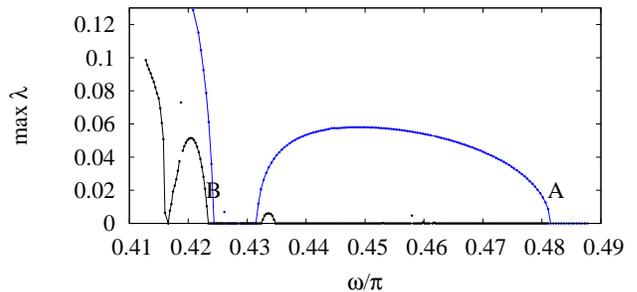}
  \caption{\label{fig:bifdiagram} Maximal Lyapunov exponents of a family of two-dimensional
    tori along a resonance channel as a function of the rotation number (on the Poincar\'e surface $\ssman$), starting from a periodic orbit
    $\ocsOa$ at $E=0.09$. Two branches of the resonance channel are
    shown; the first branch (in blue) emerges at  $\ocsOa$, the second
    branch (in black) appears at the bifurcation point of the first
    branch at $\omega/\pi=0.481422634$. The
    points of frequency halving bifurcations ``A'' and ``B'' are the bottlenecks of the transition from the trapping to roaming stage.}
\end{figure}

In the cases discussed in Sec.~\ref{sec:transition}, trajectories
undergo a transition (after a trapping stage) in the vicinity of a
nonresonant elliptic periodic point ${\bf x}_0$, whether it is
associated with $\ocsOa$ or $\ocsOc$. For each of these periodic
points, the matrix $D\mapF({\bf x}_0)$ has eigenvalues
$\exp{(\pm\iota\omega_1)}$, $\exp{(\pm\iota\omega_2)}$ (numerical
values are given in Tab.~\ref{tab:crd1}.) Processes associated with
the escape from the trapping stage can be better understood by
analyzing the tangent space of the elliptic periodic orbit ${\mathcal
O}_a$ that locally has the structure of a direct product (center $+$
center) $\torus^1\times{\mathrm I}_1\times\torus^1\times{\mathrm
I}_2$, with the periodic orbit at the origin. The elements of the two
intervals ${\mathrm I}_i\subset{\mathbb R}$ are rotation numbers
$\omega_i$, which are not unique in general: The choice is fixed by
requiring $\lim_{\epsilon \rightarrow 0}\omega_i=\omega^0_i$, where $\epsilon$
is a measure of the ``diameter'' of the torus and $\omega^0_i$ are
stability angles of the organizing periodic orbit. The Poincar\'{e}
map $\mapF$ induces rotations on $\torus^1$, $r_{\omega_1}\times{\mathrm
1}\times r_{\omega_2}\times{\mathrm 1}$, where $r_{\omega}$ is a
rotation on $\torus^1$ with the rotation number $\omega$. Partial
(or complete) resonances are determined by one (or two) resonance
conditions $n\omega_1+m\omega_2+k=0$, where $(n,m,k)$ are integers
such that $|n|+|m|+|k|>0$. The most striking trapping effects are
observed for partial resonances of the type $\torus^1\times{\mathrm
I}_1\times\{0\}\times\{0\}$, and $\{0\}\times\{0\}\times\torus^1\times{\mathrm I}_2$. They are two-dimensional manifolds (locally),
and can be foliated by one-dimensional invariant closed curves, called
hereafter ``loops.'' We propose to investigate a resonance manifold by
mapping out dynamical invariants that form its backbone structure.
Choosing either of the two situations, a resonance channel has been
constructed by finding the two-dimensional invariant tori for
$\omega_i\in{\mathrm I}_i$. At a small distance from the periodic
orbit we use information obtained from the linear normal form
$D\mapF({\bf x}_0)$.  Once an initial loop is found, we follow the
progress as the rotation number $\omega$ is varied continuously
monitoring their stability properties. Local normal stability of each
family of tori can be represented by plotting the maximal Lyapunov exponent
$\lambda$ by solving the generalized eigenvalue problem [See
Appendix~\ref{sec:appB} and Eq.~(\ref{e:Dloop})], versus the rotation number $\omega$. Such a plot for a
family of two-dimensional tori, originating from $\ocsOa$, is shown
in~\figref{fig:bifdiagram}.  From Fig.~\ref{fig:bifdiagram}, we
obtain a transition point at $\omega/\pi\approx 0.481$ in the form of a bifurcation of an
invariant torus. Projections of two-dimensional invariant tori in
the transition regions are shown in Figs.~\ref{fig:str1}
and~\ref{fig:str2}, while corresponding loops in the surface of
section $\ssman$ are shown in Figs.~\ref{fig:ps1} and~\ref{fig:ps3}.

The transition stage (see Figs.~\ref{fig:ps1} and \ref{fig:ps3})
indicates an exponential divergence resulting from an escape along the
unstable branch of a hyperbolic manifold.  The maximal Lyapunov exponent of
the segment of a trajectory in the capture stage can be estimated by
observing the duration of capture, and the per-return Lyapunov
exponent can be estimated as $\lambda\simeq 1/N$, where $N$ is the
number of returns to the surface of section $\ssman$ before the
escape. In Fig.~\ref{fig:ps1} the proximity of the torus to a 2:5
resonance zone suggests the influence of a periodic orbit with or 2:5
winding number ratio (in hollow circles). For trajectory close to
orbit $\ocsOa$ we have $N\approx 150$, yielding a typical value of
$\lambda\simeq 0.007$.  This value is inconsistent with the Lyapunov
exponent of the nearby resonant periodic orbit (which has a
Lyapunov exponent of $\lambda=0.113$ per return to the surface of
section $\ssman$), indicating that other structures than periodic orbits are
important in describing the capture processes. Unstable two-dimensional tori are indeed better candidates for the escape scenario~: An
estimate of the Lyapunov exponent in the escape stage is consistent
with the scenario of escape along unstable manifolds of the resonant
orbit.  The local rate of transition at the onset is estimated by the
largest Lyapunov exponent in the family. In the case shown
in~Fig.~\ref{fig:bifdiagram} it is close to $\lambda=0.06$. The full
picture of dynamics is complicated by existence of a family of tori
with varying (and smaller) Lyapunov exponents.

\begin{figure}[t]
  \begin{center}
    \includegraphics[width=8.2cm]{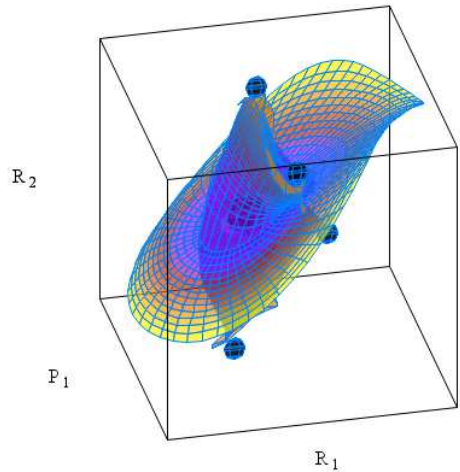}
    \caption{\label{fig:surface4d} Geometry of
      the families of two-dimensional tori of Fig.~\ref{fig:bifdiagram} (for $E=0.09$). The two branches of tori are displayed in the
      ($R_1$,$P_1$,$R_2$) projection of their Poincar\'e section (consequently one branch is composed of one-dimensional curves). The first branch (nearly horizontal and corresponding to the blue curve in Fig.~\ref{fig:bifdiagram}) emerges close
      to the fixed point $\ocsOa$ (in the center of the figure). The
      second branch (nearly vertical and corresponding to the black curve in Fig.~\ref{fig:bifdiagram}) emerges at the bifurcation point of the first branch
      (with $\omega/\pi=0.481422634$). We also represent the points (in black)
      of the 2:5 resonant periodic orbit which obstructs
      the continuation of the second branch. Note that this figure
      links Fig.~\ref{fig:ps1} (where two different projections are
      plotted and only one torus shown) with Fig.~\ref{fig:bifdiagram}.}
  \end{center}
\end{figure}

\begin{figure}
  \includegraphics[width=8.5cm]{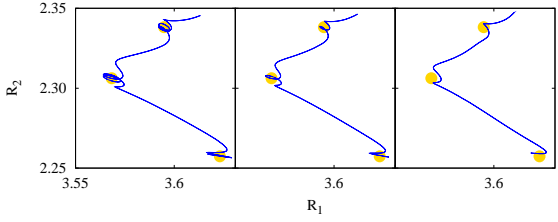}
  \caption{Approach to a rational rotation number 2:5 in the second branch
    of a family of tori of periodic orbit $\ocsOa$, $E=0.09$ (see
    also~\figref{fig:bifdiagram}.) Rotation numbers, from right to left are
    $2\omega/\pi=0.4148$,$0.4110$,$0.4104$.  Nontrivial foliation around
    manifolds of a resonant 2:5 periodic orbit can be
    seen. Note that this figure is related to frequency-halved loops
    in the second branch of~\figref{fig:bifdiagram}, therefore we have
    doubled the original rotation numbers.}\label{fig:s5A_bifurkat}
\end{figure}

In~Fig.~\ref{fig:surface4d} we represent the two families of
two-dimensional tori considered in Fig.~\ref{fig:bifdiagram} (blue and
black curves). First, the organizing periodic orbits $\ocsOa$ (center of Fig.~\ref{fig:surface4d})
and the resonance 2:5 (exterior spheres) are located. The
projections of the two families of loops in $\ssman$ are plotted in
the three dimensional space $(R_1, P_1, R_2)$. Meridians of the
surfaces are invariant under the Poincar\'{e} map (i.e.\ they are
invariant loops $\bm\gamma$).  The first family of tori (blue curve in
Fig.~\ref{fig:bifdiagram}) starts from the central periodic orbit
$\ocsOa$ and continues outwards as the rotation number decreases from
the value of $\omega^{(\ocsOa)}/\pi=0.49$ (see~\tabref{tab:crd1} and
the blue curve in Fig.~\ref{fig:bifdiagram}). The first loops of this
family have zero Lyapunov exponent (the ones with $\omega/\pi$ between
0.49 and 0.481).  At the bifurcation point ($\omega/\pi=0.481$), the
second family (black curve on Fig.~\ref{fig:bifdiagram}) branches off
of the first one and continues normally (with zero Lyapunov
exponent). The continuation of the first branch of tori is now
normally hyperbolic (see Fig.~\ref{fig:bifdiagram}) from $\omega/\pi$
between 0.481 to 0.432, while the new branch of frequency halved loops
is at first elliptically stable. The bifurcation at $\omega/\pi=0.481$
is a frequency-halving, since the emerging loop winds around the
original one twice, or in other words, has half the rotation number.
This process is very general and we expect it to occur in the vicinity
of any periodic orbit with several elliptic stability degrees of
freedom. The family of tori has singularities at some specific
rotation numbers, but the manifold can typically be continued across
them, and therefore seems to be robust. The behaviour of the second
branch of this family of tori as it approaches the rational rotation
number $\omega/\pi=2/5$ was investigated. A nontrivial foliation of
invariant loops in the vicinity of a 2:5 periodic orbit is shown
in~\figref{fig:s5A_bifurkat}. Allowed by dimensional analysis, a
possible scenario is that this family of tori is heteroclinic to the
invariant manifolds of other invariant tori, related to periodic
points in 2:5 resonance with $\ocsOa$. However, a picture of
interconnected families of tori, permeating bulk of the entire phase
space is yet to emerge.

From the numerical simulations of a large assembly of trajectories,
the following assumptions emerge~: 1) the lowest order $k=1$ resonance
controls the rates of transition from regular to chaotic dynamics, 2) the
$k=1$ resonance is a manifold that has a two-dimensional ``backbone''
manifold, in analogy with resonance manifolds of integrable
Hamiltonian systems, and 3) regular-to-chaotic transition occurs at
the point where there is a transition in the normal stability of this
manifold.  From these assumptions, the typical scenario for escape
after trapping by a weakly hyperbolic family of tori, is the following
one: First the trajectory evolves in a regular region until it finds
an exit channel (the transition stage) in the form of a manifold of
normally hyperbolic invariant two-dimensional tori, and follows along
a manifold becoming more chaotic progressively, as it visits
invariants with larger hyperbolicity (Lyapunov exponent). Eventually
it is escapes to a strongly chaotic region using the unstable
manifold of a hyperbolic periodic orbit with a large Lyapunov
exponent.

\section{Conclusions}
In contrast to collinear OCS where the phase space is roughly divided into islands and
chaotic seas, the phase space of planar OCS exhibits a complex ocean with currents, reefs and
shoals which slow down the progress toward energy equilibration. In this article, we have identified these structures  and their linear stability properties. Principal among them are two-dimensional invariant tori which occur in families and can be parametrized by their rotation numbers. These structures are organized around periodic orbits which provide the backbone to the dynamics. By trapping trajectories temporarily, they act as bottlenecks to the exploration of larger parts of phase space. Our work also makes explicit the mechanisms by which trajectories are trapped and by which they escape from the trap.

\acknowledgments
This research was partially supported by the US National Science Foundation. CC acknowledges financial support from the PICS program of the CNRS.

\appendix
\section{Discrete symmetries}
\label{sec:appA}
The time-reversibility of Hamiltonian~(\ref{e:ham}) induces
discrete symmetries which are taken into account to
uniquely define invariant points on the surface of section $\ssman$ and to
to evaluate multiplicities of periodic orbits.

Time reversibility symmetry, valid in each degree of freedom
individually, induces ``pmm'' (in crystallographic classification)
symmetry group $C_{2v}$ which acts on intrinsic coordinates $P_1$,
$P_2$, $\alpha$ and $P_{\alpha}$, while $R_1$ and $R_2$ are left
invariant. Elements of $C_{2v}$ are identity $e$, reflection
$\sigma_1$, reflection $\sigma_2$, and inversion $i$, defined as~:
\begin{eqnarray*}
&& e(P_1,P_2,\alpha,P_{\alpha})=(P_1,P_2,\alpha,P_{\alpha}),\\
&& \sigma_1(P_1,P_2,\alpha,P_{\alpha})=(P_1,P_2,2\pi-\alpha,-P_{\alpha}),\\
&& \sigma_2(P_1,P_2,\alpha,P_{\alpha})=(-P_1,-P_2,2\pi-\alpha,P_{\alpha}),\\
&& i(P_1,P_2,\alpha,P_{\alpha})=(-P_1,-P_2,\alpha,-P_{\alpha}).
\end{eqnarray*}
This discrete symmetry is useful for the method of surface of section,
because it allows to relate points ${\bf x}$ in phase space with $P_\alpha=0$ and
$\dot{P}_{\alpha}<0$, not on the surface $\ssman$,
with points $\sigma_1({\bf x})$ which are on the surface $\ssman$. 

In addition to exact discrete symmetries discussed above,  the
specific form of potential energy~(\ref{e:U}) induces an approximate
$R_1$--$R_2$ reflection symmetry as seen in~\figref{fig:equipotential}. 
Equation~(\ref{e:U}) can be written in the form of
\begin{eqnarray*}
  V &=& D_1\VMorse ( R_1; \beta_1, R^{0}_1 ) +
  D_2\VMorse( R_2; \beta_2, R^{0}_2 ) \\
  &&+D_3 \VMorse ( R_3; \beta_3, R^{0}_3 )+ V_I(R_1,R_2,R_3),
\end{eqnarray*}
where $\VMorse=[1-\exp(-\beta(R-R^0))]^2$.
Using $\bar{D}=(D_1+D_2)/2$, and $\delta D=(D_2-D_1)/2$, the potential is rewritten as $V(R_1,R_2,R_3) = U_0(R_1,R_2) + U_I(R_1,R_2,R_3)$, where
\begin{eqnarray*}
&& U_0 = \bar{D}\left(\VMorse ( R_1; \beta_1, R^{0}_1 )+\VMorse ( R_2; \beta_2, R^{0}_2 )\right),\\
&& U_I= \delta D \left(\VMorse( R_2; \beta_2, R^{0}_2 ) -
    \VMorse ( R_1; \beta_1, R^{0}_1 ) \right)\\
    && +D_3 \VMorse ( R_3; \beta_3, R^{0}_3 ) + V_I(R_1,R_2,R_3).
\end{eqnarray*}
This partition quantifies the approximate symmetry.
The non-vanishing parameters $\delta D$, $D_3$ and $A$ measure the
deviation from the exactly symmetry. For the planar OCS, these parameters are $\delta D=0.065$, $D_3=0.16$ and $A=0.2$ compared with $\bar{D}=0.348$.

With respect to linear transformations,
Morse potentials transform as 
$$
  \VMorse( aR+b; \beta, R^0) = 
  \VMorse( R; a\beta, (R^0-b)/a ).
$$
Considering the linear transformations of the coordinates $R_1$ and $R_2$ given by
$L(R_1,R_2) = ( a_1 R_2+b_1 , a_2R_1+b_2 )$, the symmetry line is
obtained by requiring that 
$$ U_0( L(R_1,R_2) ) = U_0(R_1, R_2)\,.$$
The solution is obtained in terms of parameters $\beta_1$ and
$\beta_2$, and in particular $\beta_1/\beta_2\approx 0.9$, and the symmetry is then given by the equation~:
\[ R_2 = \frac{\beta_1}{\beta_2} (R_1-R^0_1) + R^0_2. \]

In case of an exact symmetry, the symmetry line would be a natural
boundary of the elementary cell of the dynamics. All orbits could be
classified with respect to this symmetry as having a symmetric
partner, or being self symmetric, as usually. When the
symmetry is only approximate the cell boundary argument is no longer valid, but
the orbits can still be classified in this way, in particular, with
regards to their degeneracy.

\section{Methodology: determination of invariant tori and their linear stability properties}
\label{sec:appB}

We briefly summarize the method we used to
compute two dimensional invariant tori of a Hamiltonian system. We have seen that this is equivalent to determining
closed invariant curves (loops) of the Poincar\'e map on the chosen surface of
section $\ssman$. Furthermore we compute the linear stability properties of such objects. This method follows the one described in Ref.~\cite{jorba01}. 

\subsection{Determination of invariant tori}

In order to determine two-dimensional tori, we use the fact that
the type of internal dynamics on $\torus^1$ is likely to be a rotation.  
We assume that the Poincar\'e map $\mapF$ has an invariant curve with an
irrational rotation number $\omega$, and that there exists a map (at least
continuous) ${\bf x}:\torus^1\mapsto\ssman$ such that
Denjoy's theorem~\cite{katok95} states that
such a {\it rotation number} $\omega$ can be defined.
Let $C(\torus^1,\ssman)$ be the space of continuous functions from
$\torus^1$ in $\ssman$, and let us define the linear map
$T_{\omega}: C(\torus^1,\ssman)\mapsto C(\torus^1,\ssman)$ as the
translation by $\omega$, i.e.\ $(T_{\omega}{\bf x})(\theta)={\bf x}(\theta+\omega)$. We define ${\bf F}: C(\torus^1,\ssman)\mapsto C(\torus^1,\ssman)$ as
\begin{equation}
  \label{e:tori-F}
  {\bf F}({\bf x})(\theta) = \mapF({\bf x}(\theta)) - (T_{\omega}{\bf x})(\theta).
\end{equation}
It is clear that zeros of ${\bf F}$ in $C(\torus^1,\ssman)$ correspond to
(continuous) invariant curves of rotation number $\omega$. The determination of two-dimensional invariant tori follows from the search of zeros of this functional.

First we expand ${\bf x}(\theta)$ in a Fourier series with real coefficients,
\begin{equation}
  \label{e:tori-fourier}
  {\bf x}(\theta)=\frac{{\bf a}_0}{2} + \sum_{k>0}\left({\bf a}_k \cos \pi k \theta + {\bf b}_k \sin \pi k\theta\right),
\end{equation}
where ${\bf a}_k, {\bf b}_k \in \reals^n$ for $k\in\naturals$ ($n$ being the dimension of the flow) and
${\bf x}(\theta)$ is a periodic function with period 2,
i.e.\ ${\bf x}(\theta+2)={\bf x}(\theta)$. We truncate these series at a fixed value of $N$,
and determine an approximation to the $2N+1$ unknown
coefficients ${\bf a}_0$, ${\bf a}_k$, and ${\bf b}_k$ for $1\leq k \leq N$.
We construct the
discretized version of \eqnref{e:tori-eqn} by considering a mesh
of $2N+1$ points on $\torus^1$: 
\[
\theta_j = \frac{2 j}{2N+1} \quad \mbox{for } 0\leq j \leq 2N,
\]
where we notice that for numerical stability reasons, the length
of $\torus^1$ is taken as 2.
Given the Fourier coefficients ${\bf a}_k$, ${\bf b}_k$, the coordinates ${\bf x}(\theta_j)$ are expressed as linear functions of the coefficients
${\bf a}_k$, ${\bf b}_k$, i.e.\ ${\bf x}(\theta_j) \equiv {\bm \phi}( \{ {\bf a}_k\}, \{ {\bf b}_k\}, j)$, given
by \eqnref{e:tori-fourier}.
Accordingly, $\mapF({\bf x}(\theta_j))$ and \eqnref{e:tori-eqn}
can be considered as functions of the coefficients ${\bf a}_k$, ${\bf b}_k$:
\begin{eqnarray*}
  {\bf F}_j(\{ {\bf a}_k\},\{ {\bf b}_k\},\omega)&=& \mapF({\bm \phi}(\{ {\bf a}_k\},\{ {\bf b}_k\},j))\\
  && - {\bm \phi}(\{ {\bf a}_k\},\{ {\bf b}_k\},j+i(\omega)),
\end{eqnarray*}
for $0\leq j\leq 2N$ and where $i(\omega) = (2N+1)\omega/2$.
The coefficients ${\bf a}_k$, ${\bf b}_k$ are the unknowns in the above equation. 

We solve ${\bf F}=0$ using a Newton's iterative algorithm. At each iteration, it provides the corrections $\delta {\bf a}_k$ and $\delta {\bf b}_k$ to be added to the ${\bf a}_k$ and ${\bf b}_k$ obtained from the previous iteration. 
We approximate $(\delta{\bf a},\delta{\bf b})$ as a solution of the following equation:
\[
{\bf F}_j({\bf a},{\bf b},\nu) + \pdfrac{{\bf F}_j}{{\bf a}_k}\delta {\bf a}_k + \pdfrac{{\bf F}_j}{{\bf b}_k}\delta
{\bf b}_k +\pdfrac{{\bf F}_j}{\omega}\delta \omega = 0
\,,
\]
where ${\bf a}=({\bf a}_0,{\bf a}_1,\ldots,{\bf a}_N)$ and ${\bf b}=({\bf b}_1,\ldots,{\bf b}_N)$. The iteration ${\bf a}'={\bf a}+\delta{\bf a}$, ${\bf b}'={\bf b}+\delta{\bf b}$ and $\omega'=\omega+\delta\omega$ converges if the initial guess is close enough to the true solution. 
The above equation requires the inversion of the Jacobian of ${\bf F}_j$.
From the previous definitions it is clear that if ${\bf x}(\theta)$ is a
Fourier series corresponding to an invariant curve then, for any
$\varphi\in\torus^1$, ${\bf y}(\theta)\equiv {\bf x}(\theta+\varphi)$ is a
different Fourier series corresponding to the same invariant curve as
${\bf x}(\theta)$. This implies that the Jacobian of ${\bf F}_j$
around the invariant curve has, at least, a one-dimensional
kernel. To solve this problem we use the Singular Value Decomposition. Even if Newton's algorithm has converged, we cannot claim with certainty that a
smooth two-dimensional torus has been found. We have noticed that crude discretization
can wash out the details of non-smooth curves. Sometimes 
doubling the number of points in the discretization turns a convergent
case into a divergent one. In most cases the reliability of a solution is almost
certain when testing the spectrum of the solution (and the norm of its eigenvectors weighted by the frequency, penalizing high harmonics): a
smooth solution should contain a unit eigenvalue. This is why it is also important to monitor the linear stability properties of the curves we obtain numerically.

\subsection{Linear stability properties}
In addition to the determination of the location of the invariant tori, we compute their
linear stability properties to obtain information on the dynamics in
its (infinitesimal) neighborhood, i.e.\ eigenvalues and eigenvectors which give at first order an approximation to the invariant manifolds (stable, unstable and central) near the invariant curve. 
We consider the generalized eigenvalue problem which amounts to finding
($\Lambda$,${\bm \psi}$) such that 
\begin{equation}\label{e:Dloop}
  D\mapF({\bf x}(\theta)){\bm \psi}(\theta) = \Lambda{\bm
  \psi}(\theta+\omega).
\end{equation}
The eigenvalues $\Lambda$ have the following properties~\cite{jorba01}:
1) $\Lambda=1$ is an eigenvalue of Eq.~(\ref{e:Dloop}); the
corresponding eigenvector is the derivative of the loop ${\bf x}$,
2) if $\Lambda$ is an eigenvalue of Eq.~(\ref{e:Dloop}), then
$\Lambda\exp{(2\iota k\pi \omega)}$ is also an eigenvalue for any
$k\in\integers$,
3) the closure of the set of eigenvalues of Eq.~(\ref{e:Dloop}) is
a union of circles centered at the origin.

There are two unit eigenvalues in the spectrum of $D\mapF$. The symplectic symmetry implies that the tori are degenerate
in the linear approximation. It implies the existence of a family of
(smooth) two-dimensional tori.  As it is usual, we expect that this family
is discontinuous and the discontinuities are around rational rotation
numbers $\omega = m/n$.  Numerically, once an invariant torus with a
specific $\omega$ is found, we simply increment the frequency
parameter $\omega\rightarrow \omega+\delta\omega$ and restart the
search. In this way, we determine these families of two-dimensional
tori parametrized by their frequency $\omega$ on the Poincar\'e
section.  More details on the algorithm are given in
Ref.~\cite{RPthesis}.

\newcommand{\noopsort}[1]{} \newcommand{\printfirst}[2]{#1}
  \newcommand{\singleletter}[1]{#1} \newcommand{\switchargs}[2]{#2#1}

\end{document}